\documentclass[10pt]{amsart}
\usepackage{amsfonts,amsmath,amssymb,latexsym}
\usepackage{multirow}
\usepackage[longnamesfirst,round]{natbib}
\usepackage{hyperref,graphicx}
\usepackage{setspace}
\usepackage{enumitem}
\makeatletter
\providecommand{\leftleadsto}{%
  \mathrel{\mathpalette\reflect@squig\relax}%
}
\newcommand{\reflect@squig}[2]{%
  \reflectbox{$\m@th#1\leadsto$}%
}
\makeatother

\newcommand{\e}{\mathbb{E}}
\newcommand{\var}{\mathbb{V}}
\newcommand{\p}{\mathbb{P}}
\newcommand{\bs}{\boldsymbol}
\newcommand{\mbf}{\mathbf}
\newcommand{\vg}{\mbf{g}}
\newcommand{\diag}{\mathop{\mathrm{diag}}}
\newcommand{\vdiag}{\mathop{\mathrm{vecdiag}}}

\newcommand{\rank}{\mathop{\mathrm{rank}}}

\newcommand{\logit}{\mathop{\mathrm{logit}}}

\newcommand{\nrm}{\mathcal{N}}

\newcommand{\bx}{\bs{x}}
\newcommand{\loc}{\bs{s}}
\newcommand{\by}{\bs{y}}
\newcommand{\bY}{\bs{Y}}

\newcommand{\bZ}{\bs{Z}}
\newcommand{\bU}{\bs{U}}

\newcommand{\ba}{\bs{a}}

\newcommand{\bp}{\bs{p}}

\newcommand{\bv}{\bs{v}}

\newcommand{\bzero}{\bs{0}}
\newcommand{\bone}{\bs{1}}
\newcommand{\bbeta}{\bs{\beta}}
\newcommand{\bdelta}{\bs{\delta}}
\newcommand{\bgamma}{\bs{\gamma}}
\newcommand{\bEta}{\bs{\eta}}
\newcommand{\bzeta}{\bs{\zeta}}
\newcommand{\bpsi}{\bs{\psi}}
\newcommand{\bsigma}{\bs{\sigma}}
\newcommand{\btheta}{\bs{\theta}}
\newcommand{\bxi}{\bs{\xi}}
\newcommand{\bhat}{\hat{\bs{\beta}}}
\newcommand{\bmu}{\bs{\mu}}

\newcommand{\mx}{\mbf{X}}
\newcommand{\umx}{\bs{\mathcal{X}}}

\newcommand{\ma}{\mbf{A}}

\newcommand{\md}{\mbf{D}}
\newcommand{\mm}{\mbf{M}}

\newcommand{\mv}{\mbf{V}}

\newcommand{\mr}{\mbf{R}}
\newcommand{\mq}{\mbf{Q}}

\newcommand{\mF}{\mbf{F}}
\newcommand{\mlam}{\mbf{\Lambda}}
\newcommand{\msigma}{\mbf{\Sigma}}

\newcommand{\proj}{\mbf{P}}
\newcommand{\id}{\mbf{I}}
\newcommand{\reals}{\mathbb{R}}
\newcommand{\meat}{\bs{\mathcal{J}}}
\newcommand{\info}{\bs{\mathcal{I}}}

\newcommand{\cml}{\text{\textsc{cml}}}

\newcommand{\pl}{\text{\textsc{pl}}}
\newcommand{\mi}{\text{\textsc{mi}}}
\newcommand{\bsf}{\text{\textsc{bsf}}}

\def\citeapos#1{\citeauthor{#1}'s (\citeyear{#1})}

\title{Spatial Regression and the Bayesian Filter}  
\author[Hughes]{John Hughes\\
Department of Biostatistics and Informatics\\
University of Colorado Denver}

\date{}

\begin{document}

\begin{abstract}
Regression for spatially dependent outcomes poses many challenges, for inference and for computation. Non-spatial models and traditional spatial mixed-effects models each have their advantages and disadvantages, making it difficult for practitioners to determine how to carry out a spatial regression analysis. We discuss the data-generating mechanisms implicitly assumed by various popular spatial regression models, and discuss the implications of these assumptions. We propose Bayesian spatial filtering as an approximate middle way between non-spatial models and traditional spatial mixed models. We show by simulation that our Bayesian spatial filtering model has several desirable properties and hence may be a useful addition to a spatial statistician's toolkit.
\end{abstract}

\maketitle

\section{Introduction} 
\label{intro}

Spatially referenced data arise in sundry fields of inquiry, e.g., radiology, neuroscience, epidemiology, marketing, ecology, agriculture, forestry, geography, and climatology. Because spatial data tend to exhibit spatial dependence (usually attractive but sometimes repulsive or even a combination of the two), a number of statistical models, collectively referred to as spatial models, have been developed for analyzing such data \citep{ban:carl:gelf:2014}. Since dependence is customarily considered to be a second-moment phenomenon, nearly all spatial models are second-moment models. In fact, second-moment methods so dominate the field that allowing ``second-moment" to be a defining characteristic of spatial models would not be unreasonable. Here we revisit this important assumption, and discuss what the assumption implies regarding the data-generating process. Our goals are to (i) provide an appreciation of the assumptions underpinning our models, and (ii) understand how these assumptions may impact the results of a spatial regression analysis. 

Often, the aim of a spatial analysis is to do inference regarding the effects $\bbeta=(\beta_1,\dots,\beta_p)'$ of a number of spatially structured covariates $\mx=(\bx_1\,\cdots\,\bx_p)$. By accounting for spatial dependence in excess of that explained by $\mx\bbeta$, it is claimed, spatial regression models permit more reliable inference for $\bbeta$, and better prediction, than do non-spatial models. But whether a given spatial model yields improved regression inference and/or prediction depends on the posited data-generating mechanism (i.e., the ``true" model from which the data arose) as well as the properties of said spatial model.

The rest of this manuscript is organized as follows. In Section~\ref{ontology} we review the class of spatial models and discuss them as data-generating mechanisms. In Section~\ref{analytic} we discuss how our modeling assumptions impact spatial regression inference and prediction. In Section~\ref{computing} we discuss computing for spatial regression. In Section~\ref{simstudy} we apply six  regression models to simulated outcomes in an effort to assess their performance in a challenging, but realistic, setting informed by the discussion in Sections~\ref{ontology} and \ref{analytic}. We develop Bayesian spatial filtering, a new approach to spatial regression, in Section~\ref{bayesfilter}. We then conclude in Section~\ref{conclusion}.

\section{Spatial Data: Ontology versus Phenomenology}
\label{ontology}

In this section we will examine spatial models as data-generating mechanisms. We begin by reviewing the most commonly applied spatial regression models---partly to introduce useful notation, and partly to highlight the models' second-order components. Then we will discuss what sort of generating mechanism we are assuming when we apply each of these models.

\subsection{A Brief Review of Spatial Regression Models}
\label{review}

Let $\bZ=(Z_1,\dots,Z_n)'$ be the response vector, where $Z_i$ is observed at spatial location $\loc_i$. If said locations are points residing in a continuous spatial domain (e.g., a Borel subset of $\reals^2$ or near the surface of a biaxial ellipsoid), the outcomes are said to be {\em point-level} or {\em geostatistical}. If $\loc_i$ instead refers to an area over which measurements have been aggregated (e.g., county, voxel, Census tract) to produce $Z_i$, the outcomes are said to be {\em areal}.

Along with $\bZ$ we have $p$ covariates $\bx_1,\dots,\bx_p$, where $\bx_j=(x_{1j},\dots,x_{nj})'$ and $x_{ij}$, like $Z_i$, was measured at spatial location $\loc_i$. Presumably, each of $\bx_1,\dots,\bx_p$ is spatially structured and so may be useful for explaining a significant portion of the spatial variation exhibited by $\bZ$.

It is often the case that $\bZ$ exhibits additional spatial structure, i.e., spatial structure that cannot be explained by $\mx\bbeta$ alone. The most common means of accounting for this additional structure is to augment the linear predictor $\mx\bbeta$ with spatially dependent random effects. This leads to the spatial generalized linear mixed model (SGLMM), for which the transformed conditional mean vector is given by
\begin{align}
\label{sglmm}
 \vg(\bmu) &= \mx\bbeta+\bpsi,
\end{align}
where $\vg(\bmu)=\langle g(\mu_1),\dots,g(\mu_n)\rangle$, $g$ is a link function, $\mu_i=\e(Z_i\mid \psi_i)$, and $\bpsi=(\psi_1,\dots,\psi_n)'$ are latent spatially dependent random effects. Conditional on $\bpsi$, the outcomes are assumed to be independent draws from a suitable distribution (common choices are binomial, Gaussian, and Poisson). Whether the spatial domain is continuous \citep{Diggle:1998p264} or discrete \citep{Besa:York:Moll:baye:1991}, the spatial random effects are nearly always assumed to be multinormal with mean $\bzero$ \citep{hara:2009}, and so variants of the SGLMM are distinguished by alternative specifications of $\bpsi$'s covariance matrix $\msigma$, which is usually structured to accommodate (or induce) spatial clustering.

For areal data, spatial proximity is defined in terms of an undirected $n$-graph $G=(V,E)$, where $V=\{1,\dots,n\}$ are the vertices and $E\subset V\times V$ are the edges. The vertices of $G$ represent the areal units, and the edges of $G$ represent adjacencies among the units (usually, a pair of vertices share an edge iff their corresponding areal units share a boundary). In this setting $\msigma$ is typically a function of $G$'s adjacency matrix---$\ma=(\ma_{uv}=1\{(u,v)\in E\})$---and perhaps one or more dependence parameters. A famous possibility is the proper conditional autoregressive (CAR) model, in which $\msigma$ is equal to $(\tau\mq)^{-1}$, where $\tau>0$ is a smoothing parameter and $\mq=\diag(\ma\bone)-\rho\ma$, with $\rho\in[0,1)$ behaving like a range parameter. This implies that $\bpsi$ is a Gaussian Markov random field (GMRF) \citep{GMRFbook}, which implies that $\psi_u$ and $\psi_v$ are independent conditional on their neighbors iff areal units $u$ and $v$ are not adjacent. That $G$'s adjacency structure corresponds to a conditional independency structure for $\bpsi$ is widely considered to be an appealing characteristic of this and similar definitions of $\msigma$. Unfortunately, the resulting marginal dependence structure for $\bpsi$ may be counterintuitive or even pathological \citep{wall2004close,Assuncao:2009p916}.

For point-level observations, the elements of $\msigma$ are given by a spatial covariance function: $\msigma_{uv}=k(\loc_u,\loc_v)$. A common choice for $k$ is the M\'{a}tern covariance function, which is given by
\begin{align*}
k(\loc_u,\loc_v) &= k_{\sigma,\nu,\rho}(\Vert\loc_u-\loc_v\Vert)=\sigma^2\frac{2^{1-\nu}}{\Gamma(\nu)}\left(\frac{\sqrt{2\nu}\Vert\loc_u-\loc_v\Vert}{\rho}\right)^\nu K_\nu\left(\frac{\sqrt{2\nu}\Vert\loc_u-\loc_v\Vert}{\rho}\right),
\end{align*}
where $\Vert\loc_u-\loc_v\Vert$ is the distance between $\loc_u$ and $\loc_v$, $\sigma^2$ is the common variance, $\nu>0$ is a smoothness parameter, $\Gamma$ denotes the gamma function, $\rho>0$ is a range parameter (often referred to as the {\em characteristic length scale}), and $K_\nu$ is the modified Bessel function of the second kind. This defines a Gaussian process \citep{rasmussen2006gpm}. Since $k$ depends only on distances between locations, the process is stationary, i.e., translation invariant. If the norm is the Euclidean norm, the process is also isotropic, which is to say that the variability is the same in all directions.

A second approach to accommodating/inducing extra-$\mx\bbeta$ spatial structure in areal outcomes is to augment $\mx\bbeta$ with an autocovariate in place of the SGLMM's random effects, in which case the linear predictor is given by
\begin{align}
\label{automodel}
\vg(\bmu) &= \mx\bbeta+\kappa\ma\{\bZ-\vg^{-1}(\mx\bbeta)\},
\end{align}
where $\mu_i=\e(Z_i\mid\{Z_j:(i,j)\in E\})$ and dependence parameter $\kappa$ captures the ``reactivity" of the outcomes to their neighbors, conditional on the independence expectations $\e(\bZ\mid\kappa=0)=\vg^{-1}(\mx\bbeta)$. (A positive value of $\kappa$ implies spatial attraction while a negative value implies repulsion, and larger $\vert\kappa\vert$ produces/indicates stronger dependence.) This defines the automodel \citep{Besa:spat:1974}, a type of Markov random field (MRF) model \citep{kindermann1980markov,Clif:mark:1990}. The proper CAR model described above is a special case. Another noteworthy example is the autologistic model \citep{Caragea:2009p778,Hughes:2011p1280} for binary data, for which (\ref{automodel}) takes the form
\begin{align*}
\logit(\bmu) &= \mx\bbeta+\kappa\ma\{\bZ-\bzeta\},
\end{align*}
where
\[
\bzeta=\{\bone + \exp(-\mx\bbeta)\}^{-1},
\]
or, more explicitly,
\begin{align*}
\log\frac{\p(Z_i=1\mid\{Z_j:(i,j)\in E\})}{\p(Z_i=0\mid\{Z_j:(i,j)\in E\})} &= \bx_i'\bbeta+\kappa\sum_{j:(i,j)\in E}[Z_j-\{1+\exp(-\bx_j'\bbeta)\}^{-1}],
\end{align*}
for $i=1,\dots,n$.

A third, and newer, type of spatial regression model is the spatial copula regression model (SCRM) \citep{Kazianka:2010p941,hughes2014copcar}. Unlike the SGLMM and automodel, the SCRM is a marginal model, which is to say the regression coefficients have the same interpretation as in the classical GLM \citep{McCu:Neld:gene:1983}. A common choice for the joint component of the spatial CRM is the spatial Gaussian copula
\begin{align*}
 \Phi_{\bzero,\mr}\{\Phi^{-1}(u_1),\dots,\Phi^{-1}(u_n)\},
\end{align*}
where the $u_i$ are standard uniform, $\Phi_{\bzero,\mr}$ denotes the cdf of the multinormal distribution with mean vector $\bzero$ and spatial correlation matrix $\mr$, and $\Phi^{-1}$ is the standard normal quantile function. See \citet{joecopulabook} for an extensive treatment of copula models, and \citet{Kolev:2009p947} for a review of copula-based regression models.

The copula can be applied to the outcomes directly, or be employed in a hierarchical fashion. The gamma--Poisson model provides an intuitive example of the latter:
\begin{align*}
Z_i\mid\lambda_i & \;\;\stackrel{\text{ind}}{\sim}\;\; \mathcal{P}\left(\lambda_i\right)\\
\lambda_i & \,\;\;\sim\;\;\, \mathcal{G}\left(\nu \mu_i,\,\nu\right)\\
\{\psi_i=\Phi^{-1}\left\{F_i\left(\lambda_i\right)\right\}\}_{i=1}^n & \,\;\;\sim\;\;\, \nrm\left(\bzero,\,\mr\right),
\end{align*}
where $\mathcal{P}$ denotes the Poisson distribution, $\mathcal{G}$ denotes the gamma distribution, $\mu_i=g^{-1}(\bx_i'\,\bbeta)$, and $F_i$ is the $\mathcal{G}(\nu\mu_i,\,\nu)$ cdf. In this formulation the copula is applied to the $\lambda_i$ (which are marginally gamma and exhibit Gaussian dependence), and so the outcomes are dependent because the $\lambda_i$ are dependent.

Two additional spatial regression models are the simultaneous autoregressive model \citep{cres:1993} and the clipped random field \citep{DeO:baye:2000}. Although interesting, these models are not applied as often as the models described above, and so, in the interest of brevity, we will not consider them further in this work.

\subsection{Interpreting Spatial Regression Models}
\label{meaning}

What do the above mentioned models---the SGLMM, the automodel, and the SCRM---{\em mean} if we attempt to grant ontological status to their second-order components? This is clearly not an issue for $\mx$  since we are in possession of it and believe it to be more fundamental than the outcomes (in the sense that much of the spatial variation exhibited by the response can be attributed to $\mx$). What we seek are equally fundamental interpretations of the models' dependence components.

Let us first consider the SGLMM, which induces extra-$\mx\bbeta$ spatial variation by augmenting the classical linear predictor with spatially dependent random effects $\bpsi$. To what aspect of reality does $\bpsi$ refer? A {\em prima facie} interpretation of $\bpsi$ would lead us to conclude that $\bpsi$ is an unobservable realization of some spatial process (just as each column of $\mx$ is an observable realization of some spatial process) and that said process acts on the outcomes on link scale and in an additive fashion. But this interpretation of $\bpsi$ does not {\em explain} the extra-$\mx\bbeta$ spatial variation in the outcomes. This interpretation merely {\em accommodates}, i.e., reveals the pattern of, that additional variation but cannot describe its origin. That is, this apparently ontological interpretation of $\bpsi$ is, in fact, phenomenological---is, in fact, no more fundamental than the outcomes themselves.

It is perhaps just as difficult to tie the automodel's autocovariate term $\kappa\ma\{\bZ-\vg^{-1}(\mx\bbeta)\}$ to (non-mathematical) reality. Since the autocovariate, unlike $\bpsi$, involves $\mx\bbeta$, one might argue that the autocovariate is more fundamental than $\bpsi$. But the autocovariate is also self-referential, i.e., it contains the response we aim to explain. And so it is not clear how one might arrive at a sensible realist interpretation of the autocovariate term. The term does admit an intuitive phenomenological interpretation, however: for the automodel, extra-$\mx\bbeta$ spatial variation is defined, quite explicitly, as localized departures from the independence expectations $\vg^{-1}(\mx\bbeta)$. We might attach this same interpretation to the SGLMM, although there the mechanism of departure from the independence expectations is less explicit and is not self-referential.

The copula-based model, whether the copula is applied directly or hierarchically, is a rather different sort of model since it does not induce/accommodate extra-$\mx\bbeta$ spatial variation on the scale of the link function. Instead, the copula acts by way of quantile transformations. To see this, consider the stochastic form of the copula model, where we apply the copula to the outcomes (in contrast to the hierarchical formulation given above):
\begin{align*}
\{\psi_i\}_{i=1}^n & \,\;\;\sim\;\;\, \nrm\left(\bzero,\,\mr\right)\\
U_i = \Phi(\psi_i) &\,\;\;\sim\;\;\, \mathcal{U}(0,1)\\
Z_i=F_i^{-1}(U_i) &\,\;\;\sim\;\;\, \mathcal{P}\left(\lambda_i\right),
\end{align*}
where $F_i^{-1}$ is the quantile function of the Poisson distribution with mean $\lambda_i=g^{-1}(\bx_i'\bbeta)$. Here, extra-$\mx\bbeta$ spatial variation originates in the $\psi_i$, carries over to the $U_i$ (which are marginally standard uniform and exhibit Gaussian dependence), and finally influences the outcomes through the quantile transformations $F_i^{-1}$ (which also incorporate $\mx\bbeta$). That is, the copula does not induce extra-$\mx\bbeta$ variation by additively perturbing $\mx\bbeta$ (or perturbing the $\lambda_i$ in any fashion) but instead pushes the $Z_i$ away from the $\lambda_i$ by inducing a spatial pattern among the $U_i$.

Does the copula represent some real-world mechanism? The answer must be no since $\bpsi$ in the copula model serves precisely the same role, conceptually, as does $\bpsi$ in the SGLMM. Both models can be viewed as latent Gaussian models, and what distinguishes them is merely the way in which the latent Gaussian random variable $\bpsi$ obscures $\vg^{-1}(\mx\bbeta)$.

And so it appears that the dependence components of commonly applied spatial regression models do not lend themselves to realist interpretations but are instead merely instrumental. The dependence components of these models may be capable of generating what we have termed extra-$\mx\bbeta$ spatial variation, but the models are unable to explain spatial variation in the response in the same sense that $\mx\bbeta$ can.

\subsection{Extra-$\mx\bbeta$ Spatial Variation as the Result of Model Underspecification}
\label{unmeasured}

Model underspecification offers a plausible realist explanation for extra-$\mx\bbeta$ spatial variation. Specifically, we might suppose that
\begin{align}
\label{realmod}
\vg(\bmu) &= \mx\bbeta + \umx\bgamma,
\end{align}
where the columns of $\umx$ are unmeasured spatial predictors, $\bgamma$ their effects. This implies that extra-$\mx\bbeta$ spatial variation is a first-moment phenomenon, i.e., the spatial dependence among the outcomes is due entirely to spatial structure among the predictors $\mx$ and $\umx$. This view demystifies the spatial regression problem and allows us to analyze the problem using intuitive and well-understood ideas regarding ordinary regression modeling (i.e., regression modeling for independent outcomes).

\section{Spatial Regression Models as Data-Analytic Tools}
\label{analytic}

In the setting of ordinary regression, consider four possibilities for a given model:
\begin{enumerate}[label=(\Alph*)]
\item the model is correct;
\item the model is underspecified, i.e., one or more important predictors is missing;
\item the model is overspecified, i.e., one or more predictors is redundant; or
\item the model contains extraneous predictors, i.e., one or more predictors is not related to the response or to any other predictor.
\end{enumerate}
If the true model is linear with spherical Gaussian errors, say,
\begin{enumerate}[label=(\Alph*)]
\item permits unbiased estimation of the regression coefficients and unbiased prediction, and yields accurate standard errors;
\item permits unbiased estimation of $\bbeta$ only if $\umx$ is not correlated with $\mx$, and leads to biased prediction and inflated standard errors;
\item permits unbiased estimation of the regression coefficients and unbiased prediction, but standard errors may be inflated dramatically due to collinearity; and
\item permits unbiased estimation of the regression coefficients and unbiased prediction, but standard errors may be inflated dramatically if the number of extraneous predictors is large.
\end{enumerate}
We mentioned in Section~\ref{intro} that employing a spatial regression model to account for extra-$\mx\bbeta$ spatial variation in the response can allegedly permit more reliable inference for $\bbeta$ than a non-spatial model can. Assuming (\ref{realmod}), and in light of (B), this claim implies that some spatial model(s) can remedy the absence of $\umx$, resulting in (i) more accurate estimation of $\bbeta$, better (ii) coverage and (iii) type II error rates, and (iv) more accurate prediction. Can any spatial regression model accomplish all of these tasks? That is, if the data-generating mechanism is (\ref{realmod}), can any spatial regression model, when employed not as data-generating mechanism but as data-analytic tool, accomplish (i--iv)?

Regarding (i), estimation of $\bbeta$ will be biased, perhaps badly so, unless the unmeasured predictors $\umx$ are not correlated with the measured predictors $\mx$. Some spatial models may be able to provide a surrogate for $\umx\bgamma$, but that is not the same as revealing $\umx$, for it is the relationship between $\mx$ and $\umx$, not the structure of $\umx\bgamma$, that matters when estimating $\bbeta$. In other words, no spatial model can remedy unmeasured confounding.

The absence of $\umx$ need not lead to poor prediction, however. Recall that the SGLMM and the automodel augment $\mx\bbeta$ with, respectively, spatial random effects $\bpsi$ or the autocovariate $\kappa\ma\{\bZ-\vg^{-1}(\mx\bbeta)\}$. Presumably, each of these terms aids prediction by acting as a surrogate for $\umx\bgamma$. The SCRM (in the form described above, at least) does not augment $\mx\bbeta$, and so we should expect that model to offer poorer predictive performance than the SGLMM and automodel.

Although the SGLMM offers better prediction than a non-spatial model or a copula-based model, the improvement is costly. To see this, it will prove useful to rewrite the SGLMM's linear predictor as
\begin{align}
\label{etaconf}
\vg(\bmu) &= \mx\bbeta+\proj_x\bpsi+(\id-\proj_x)\bpsi,
\end{align}
where $\proj_x=\mx(\mx'\mx)^{-1}\mx'$ is the orthogonal projection onto $C(\mx)$, and $\id$ denotes the $n\times n$ identity matrix. This form of the linear predictor allows us to see that the SGLMM is overspecified as well as underspecified: since $C(\proj_x)=C(\mx)$, the model is perfectly collinear. This trait of the SGLMM---which inflates the variance of $\bhat$, often dramatically, as per (C) above---is called {\em spatial confounding} \citep{CLAYTON:1993p1156,Reich:2006p1157,paciorek2010importance,hodges2010adding}.

The confounding evident in (\ref{etaconf}) can be eliminated by removing $\proj_x\bpsi$, thereby constraining smoothing to the residual space $C(\mx)^\perp$. This technique is called restricted spatial regression (RSR) \citep{hodges2010adding}. RSR not only obviates spatial confounding but can also permit considerable dimension reduction and much more time- and space-efficient computation \citep{sglmm,ngspatial}.

\citet{hanks} acknowledged the potential computational benefits of RSR but cautioned that RSR may lead to erroneous inference for $\bbeta$ if (\ref{sglmm}) is the true model. According to \citet{hanks}, the RSR model, which has linear predictor
\begin{align}
\label{rsr}
\vg(\bmu) &= \mx\bbeta+(\id-\proj_x)\bpsi,
\end{align}
implicitly assumes that all variation in the direction of $\mx$ can be explained by $\mx\bbeta$, whereas the traditional SGLMM can accommodate additional variation in the direction of $\mx$.

To support the latter claim they rewrite (\ref{etaconf}) as
\begin{align}
\label{hankseta}
\vg(\bmu) &= \mx\bbeta+\proj_x\bpsi+(\id-\proj_x)\bpsi\\
\nonumber&= \mx\bbeta+\mx(\mx'\mx)^{-1}\mx'\bpsi+(\id-\proj_x)\bpsi\\
\nonumber &= \mx\left\{\bbeta+(\mx'\mx)^{-1}\mx'\bpsi\right\}+(\id-\proj_x)\bpsi\\
\nonumber\dagger &= \mx\bdelta+(\id-\proj_x)\bpsi.
\end{align}
Similarly, we can rewrite our posited data-generating model (\ref{realmod}) as
\begin{align}
\label{realeta}
\vg(\bmu) &= \mx\bbeta+\proj_x\umx\bgamma+(\id-\proj_x)\umx\bgamma\\
\nonumber&= \mx\bbeta+\mx(\mx'\mx)^{-1}\mx'\umx\bgamma+(\id-\proj_x)\umx\bgamma\\
\nonumber &= \mx\left\{\bbeta+(\mx'\mx)^{-1}\mx'\umx\bgamma\right\}+(\id-\proj_x)\umx\bgamma\\
\nonumber\ddagger &= \mx\bdelta + (\id-\proj_x)\umx\bgamma
\end{align}
to show that (\ref{realmod}) can generate additional variation in the direction of $\mx$. Hence, (\ref{hankseta}) and (\ref{realeta}) show that the RSR model---$\dagger$ and $\ddagger$---can, in fact must, accommodate extra variation in the direction of $\mx$. That is, when we fit an RSR model, we are estimating $\bdelta$, not $\bbeta$, and this is true whether the ``true" linear predictor is $\mx\bbeta+\bpsi$ or $\mx\bbeta+\umx\bgamma$ (assuming $\umx$ is correlated with $\mx$).

In any case, the crux of the matter is the absence of $\umx$. It is the absence of $\umx$ that prevents accurate estimation of $\bbeta$ (if $\umx$ is correlated with $\mx$), and neither the traditional SGLMM nor the RSR model provides a remedy. What both models {\em do} provide is more accurate prediction (by furnishing a stand-in for $\umx\bgamma$). The traditional SGLMM accomplishes this at the cost of spatial confounding and a large (with respect to both time and storage) computational burden. RSR successfully addresses these problems and, if applied properly, yields significantly better predictive performance than the traditional model (see Section~\ref{simstudy} below).

Although unmeasured spatial confounding cannot be remedied (in general, or entirely, at least), \citet{hanks} suggest another avenue by which inference for $\bbeta$ might be improved. They note that the RSR model may suffer from a low coverage rate for $\bbeta$, and they recommend the larger credible region that results from posterior predictive inference \citep{Gelman2013Baysian-Data-An} according to
\[
\tilde{\bbeta}^{(k)} \sim \nrm\{\bdelta^{(k)}, (\mx'\mx)^{-1}\mx'\msigma^{(k)}\mx(\mx'\mx)^{-1}\},
\]
where $\bdelta^{(k)}$ is the $k$th sample from $\bdelta$'s posterior, and $\msigma^{(k)}=\msigma(\bxi^{(k)})$ is the value of $\msigma$ produced from the $k$th update of the covariance parameters $\bxi^{(k)}$. In Section~\ref{simstudy} we study how this approach performs in practice.

The spatial confounding caused by adding $\bpsi$ to $\mx\bbeta$ may lead us to suspect that the automodel, which adds $\kappa\ma\{\bZ-\vg^{-1}(\mx\bbeta)\}$ to $\mx\bbeta$, is likewise confounded. This is, in fact, the case for the traditional automodel, which has linear predictor
\begin{align*}
\vg(\bmu) &= \mx\bbeta+\kappa\ma\bZ.
\end{align*}
\citet{Caragea:2009p778} studied this problem in the context of the autologistic model and showed that centering the autocovariate alleviates spatial confounding for the automodel: $\ma\{\bZ-\vg^{-1}(\mx\bbeta)\}$ is to $\ma\bZ$ as $(\id-\proj_x)\bpsi$ is to $\bpsi$.

Since the SCRM does not augment $\mx\bbeta$, the SCRM is not spatially confounded. But the SCRM has no way of fitting extra-$\mx\bbeta$ spatial variation, and so we should expect the SCRM's predictive performance to be no better than that of the ordinary GLM.

\section{Some Computational Aspects of Spatial Regression}
\label{computing}

Now we turn our attention to computational issues involved in spatial regression. This topic could easily fill a book, and so our goal is not to provide a thorough treatment. We aim to describe only the most important aspects of computing for spatial regression, and, in so doing, to set the stage for the simulation study that is the subject of Section~\ref{simstudy}. We will focus on models for binary areal data, for four reasons: (1) binary spatial data are common; (2) binary outcomes, being relatively uninformative, present the most challenging case; (3) the automodel is an areal model; and (4) although spatial counts are common, the auto-Poisson and autonegative binomial models permit only negative spatial dependence. (This limitation of the auto-Poisson and autonegative binomial models can be overcome through Winzorization \citep{Kais:Cres:mode:1997}, but the resulting models are, perhaps surprisingly, not often applied.)

\subsection{Computing for the Autologistic Model}
\label{autocompute}

Maximum likelihood and Bayesian inference for the autologistic model are complicated by an intractable normalizing function. To see this, assume the underlying graph has clique number 2, in which case the joint pmf of the centered model is
\begin{align*}
\pi(\bZ\mid\btheta) = c(\btheta)^{-1}\exp\left(\bZ^\prime \mx \bbeta - \kappa \bZ^\prime \ma \bzeta + \frac{\kappa}{2}\bZ^\prime \ma \bZ\right),
\end{align*}
where $\btheta=(\bbeta',\kappa)'$ and
\[
c(\btheta)=\sum_{\bY\in\{0,1\}^n}\exp\left(\bY^\prime \mx \bbeta - \kappa \bY^\prime \ma \bzeta + \frac{\kappa}{2}\bY^\prime \ma \bY\right)
\]
is the normalizing function \citep{Hughes:2011p1280}. The normalizing function is intractable for all but the smallest datasets because the sample space $\{0,1\}^n$ contains $2^n$ points.

There are many techniques for doing inference in the presence of intractable normalizing functions \citep[see, e.g.,][]{intractable}. One way is to avoid the normalizing function altogether. For the autologistic model, this can be accomplished by considering the so called pseudolikelihood (PL), which is a composite likelihood \citep{Lindsay:1988p1155} of the conditional type. Each of the $n$ factors in the pseudolikelihood is the likelihood of a single observation, conditional on said observation's neighbors:
\begin{align*}
p_i(\btheta)^{z_i}\{1-p_i(\btheta)\}^{1-z_i} &= \p(Z_i=z_i\mid\{Z_j:(i,j)\in E\})\\
&=\frac{\exp[z_i\{\bx_i^\prime\bbeta+\kappa\ba_i'(\bZ-\bzeta)\}]}{1+\exp\{\bx_i^\prime\bbeta+\kappa\ba_i'(\bZ-\bzeta)\}},
\end{align*}
where $z_i$ is the observed value of $Z_i$, and $\ba_i'$ is the $i$th row of $\ma$. Since the $p_i$ are free of the normalizing function, so is the log pseudolikelihood, which is given by
\begin{align}
\label{logpl}
\ell_\pl(\btheta) &= \bZ^\prime\{\mx\bbeta+\kappa\ma(\bZ-\bzeta)\}-\sum_i\log[1+\exp\{\bx_i^\prime\bbeta+\kappa\ba_i'(\bZ-\bzeta)\}].
\end{align}
Although (\ref{logpl}) is not the true log likelihood unless $\kappa=0$, \citet{Besa:stat:1975} showed that the maximum pseudolikelihood estimator (MPLE) converges almost surely to the maximum likelihood estimator (MLE) as the lattice size goes to $\infty$ (under an infill, as opposed to increasing domain, regime). For small samples the MPLE is less precise than the MLE (and the Bayes estimator), but point estimation of $\bbeta$ is generally so poor for small samples that precision is unimportant. When the sample size is large enough to permit accurate estimation of $\bbeta$, the MPLE is nearly as precise as the MLE \citep{Hughes:2011p1280}.

We find the MPLE $\tilde{\btheta}$ by optimizing $\ell_\pl(\btheta)$. This is computationally efficient even for larger samples. To speed computation even further, we can use a quasi-Newton \citep{byrd1995limited} or conjugate-gradient algorithm and supply the score function
\[
\nabla\ell_\pl(\btheta)=((\bZ-\bp)^\prime(\id-\kappa\ma\md)\mx, (\bZ-\bp)^\prime\ma(\bZ-\bzeta))^\prime,
\]
where $\bp=(p_1,\dots,p_n)^\prime$ and $\md=\diag\{\zeta_i(1-\zeta_i)\}$.

Confidence intervals can be obtained using a parametric bootstrap \citep{efron1994introduction} or sandwich estimation. For the former we generate $b$ samples from $\pi(\bZ\mid\tilde{\btheta})$ and compute the MPLE for each sample, thus obtaining the bootstrap sample $\tilde{\btheta}^{(1)},\ldots,\tilde{\btheta}^{(b)}$. Appropriate quantiles of the bootstrap sample are then used to construct approximate confidence intervals for the elements of $\btheta$.

The second approach for computing confidence intervals is based on \citep{varin2011overview}
\begin{align}
\label{sandwich}
\sqrt{n}(\tilde{\btheta}-\btheta)\;\Rightarrow\;\nrm\{\bzero,\,\info_\pl^{-1}(\btheta)\meat_\pl(\btheta)\info_\pl^{-1}(\btheta)\},
\end{align}
where $\info_\pl^{-1}(\btheta)\meat_\pl(\btheta)\info_\pl^{-1}(\btheta)$ is the Godambe information matrix \citep{godambe1960optimum}. The ``bread" in this sandwich is the inverse of the information matrix $\info_\pl(\btheta)=-\e\nabla^2\ell_\pl(\btheta)$, and the ``filling" is the variance of the score: $\meat_\pl(\btheta)=\e\nabla\nabla^\prime\ell_\pl(\btheta)$. We use the observed information (computed during optimization) in place of $\info_\pl$ and estimate $\meat_\pl$ using a parametric bootstrap. For the bootstrap we simulate $b$ samples $\bZ^{(1)},\dots,\bZ^{(b)}$ from $\pi(\bZ\mid\tilde{\btheta})$ and estimate $\meat_\pl$ as
\[
\hat{\meat}_\pl(\tilde{\btheta})=\frac{1}{b}\sum_{k=1}^b\nabla\nabla^\prime\ell_\pl(\tilde{\btheta}\mid\bZ^{(k)}).
\]

Because the bootstrap sample can be generated in parallel and little subsequent processing is required, these approaches to inference are very efficient computationally, even for large datasets. We note that sandwich estimation tends to be much faster than the full bootstrap. Moreover, asymptotic inference and bootstrap inference yield comparable results for practically all sample sizes because (\ref{sandwich}) is not, in fact, an asymptotic result. This is because the log pseudolikelihood is approximately quadratic with Hessian approximately invariant in law, which implies that the MPLE is approximately normally distributed irrespective of sample size \citep{Geyer2005Le-Cam-Made-Sim}.

\subsection{Computing for the Traditional SGLMM}
\label{sglmmcompute}

The traditional SGLMM is typically applied using MCMC for Bayesian inference, in which case the model for $\bpsi$ might be considered a prior distribution. Whether the model is viewed from a Bayesian or a classical point of view, or is applied to areal data or point-level data, the computational bottleneck is the handling of $\bpsi$'s precision matrix $\msigma^{-1}$.

For point-level outcomes the customary approach to this problem is to avoid inversion of $\msigma$ in favor of Cholesky decomposition followed by a linear solve. Since $\msigma$ is typically dense, its Cholesky decomposition is in $O(n^3)$, and so the time complexity of the overall fitting algorithm is in $O(n^3)$. This considerable computational expense makes the analyses of large point-level datasets time consuming or infeasible. Consequently, efforts to reduce the computational burden have resulted in an extensive literature detailing many approaches, e.g., process convolution \citep{Higdon:2002p909}, fixed-rank kriging \citep{Cressie:2008p911}, Gaussian predictive process models \citep{Banerjee:2008p903}, covariance tapering \citep{Furrer:2006p912}, approximation by a Gaussian Markov random field \citep{Rue:2002p915,lindgren2011explicit}, integrated nested Laplace approximations \citep{rue2009approximate}, and nearest-neighbor Gaussian process models \citep{datta2016hierarchical}.

Fitting the areal version of the model can also be burdensome even though the areal model is parameterized in terms of $\msigma^{-1}$ and $\msigma^{-1}$ is sparse. It is well known that a univariate Metropolis--Hastings algorithm for sampling from the posterior distribution of $\bpsi$ leads to a slow mixing Markov chain because the components of $\bpsi$ exhibit strong \emph{a posteriori} dependence. This has led to a number of methods for updating the random effects in a block(s). Constructing proposals for these block updates is challenging, and the improved mixing comes at the cost of increased running time per iteration \citep[see, for instance,][]{Knor:Rue:on:2002,Haran:2003p921,Haran:2010p922}.

The large dimension of $\bpsi$ and the slowness of mixing together imply a large storage requirement too. If RAM capacity is insufficient the samples can be stored in a file-backed structure, but this solution is hardly ideal since accessing secondary storage is many orders of magnitude slower than accessing RAM.

\subsection{Computing for the RSR Model}
\label{rsrcompute}

Restricted spatial regression can be done parsimoniously and efficiently by augmenting $\mx\bbeta$ with an appropriate basis expansion. \citet{sglmm} employed the linear predictor $\mx\bbeta+\mm\bEta$ in their sparse areal mixed model (SAMM), where $\mm$ is $n\times q$ and its columns are the $q$ principle eigenvectors of the Moran basis; $\bEta\sim\nrm\{\bzero,(\tau\mm'\mq\mm)^{-1}\}$ are spatial random effects; and $\mq=\diag(\ma\bone)-\ma$ is the Laplacian \citep{graphspectra} of $G$. The Moran basis takes its name from the Moran operator for $\mx$: $M_x=(\id-\proj_x)\ma(\id-\proj_x)$. This operator appears in a generalized form of Moran's $I$ (a popular nonparametric measure of spatial dependence for areal data \citep{Moran:1950p874}), which is given by
\[
I_x(\bv)=\frac{n}{\bone'\ma\bone}\frac{\bv'(\id-\proj_x)\ma(\id-\proj_x)\bv}{\bv'(\id-\proj_x)(\id-\proj_x)\bv}.
\]
(This becomes Moran's $I$ when $\proj_x$ is replaced with $n^{-1}\bone\bone'$, i.e., when $\mx=\bone$.)

\begin{figure}[h]
   \centering
   \begin{tabular}{ccc}
   \includegraphics[scale=.3]{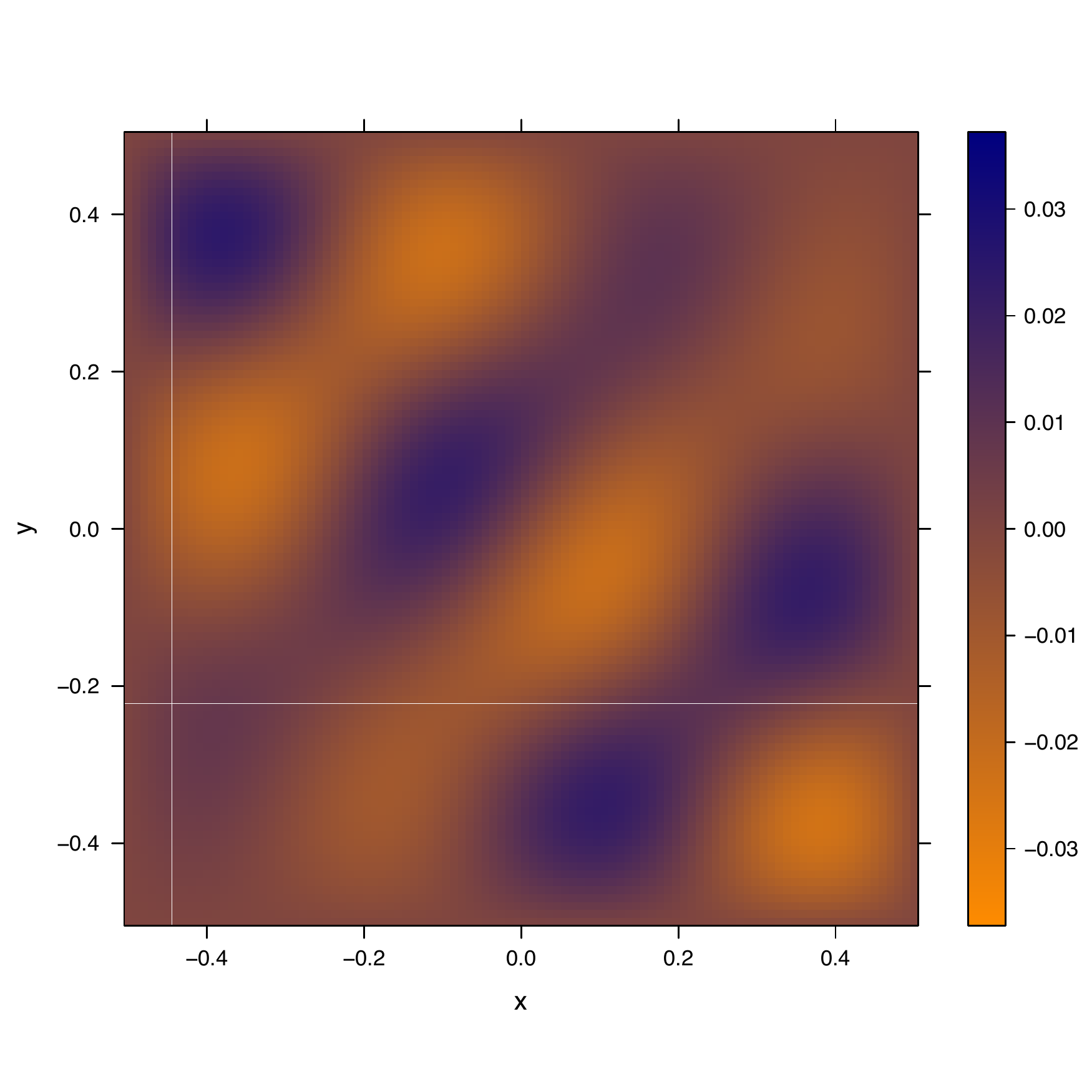} & \includegraphics[scale=.3]{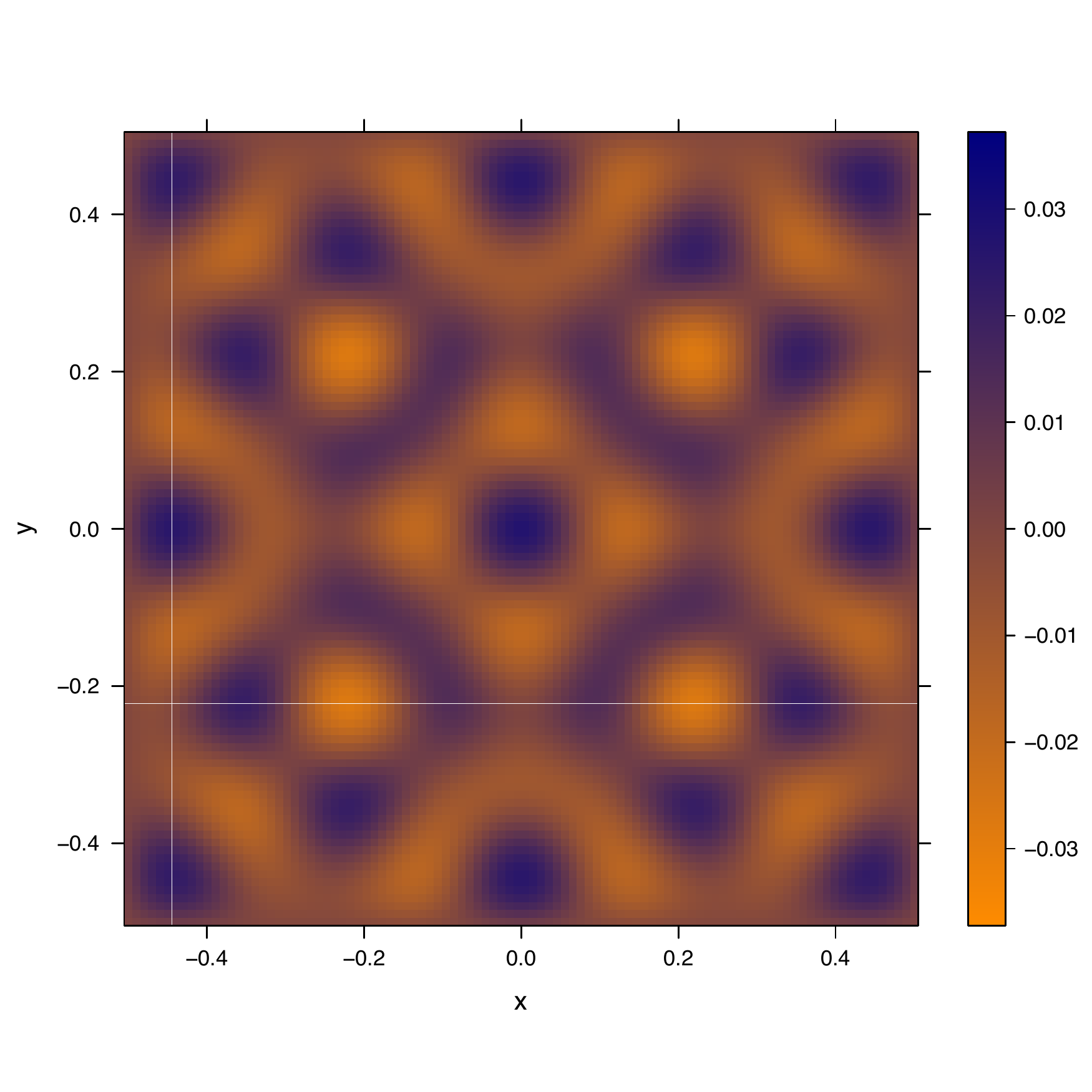} & \includegraphics[scale=.3]{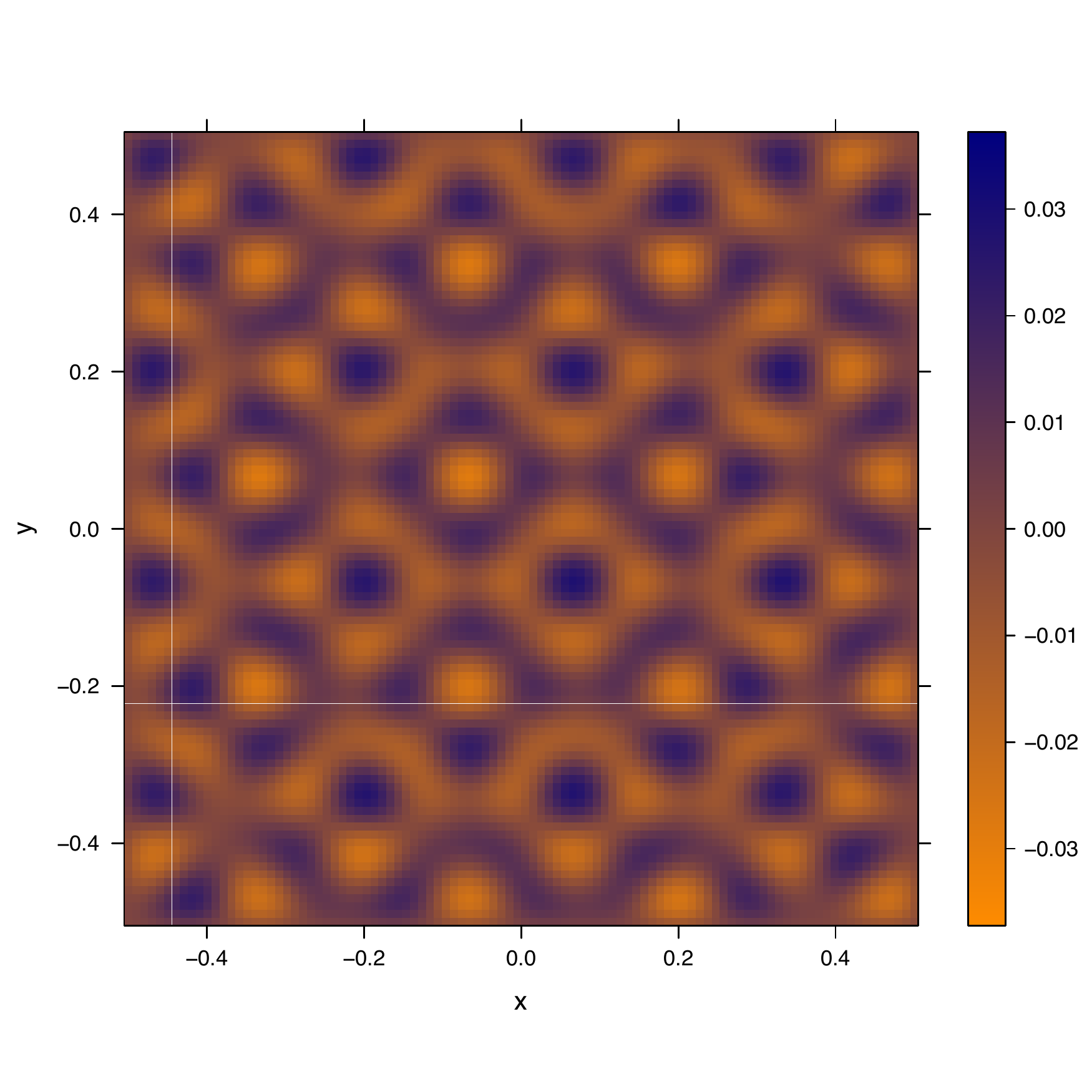}
   \end{tabular}
   \caption{Three Moran basis vectors, exhibiting spatial patterns of increasingly finer scale.}
   \label{moranfig}
\end{figure}

\citet{Boots:2000p914} showed that (1) the (standardized) spectrum of $M_x$ comprises the possible values for $I_x$, and (2) the eigenvectors comprise all possible mutually distinct patterns of clustering residual to $C(\mx)$ and accounting for $G$. The positive (negative) eigenvalues of $M_x$ correspond to varying degrees of positive (negative) spatial dependence, and the eigenvectors associated with a given eigenvalue ($\omega_i$, say) are the patterns of spatial clustering that data exhibit when the dependence among them is of degree $\omega_i$. In other words, the eigenvectors of $M_x$ form a multiresolutional spatial basis for $C(\mx)^\perp$ that exhausts all possible patterns that can arise on $G$. Three Moran basis vectors are shown in Figure~\ref{moranfig}.

Since we do not expect to observe repulsion in the phenomena to which these models are usually applied, we can use the spectrum of the operator to discard all repulsive patterns, retaining only attractive patterns for our analysis (although it can be advantageous to accommodate repulsion \citep{Griffith:2006p1579}). By retaining only eigenvectors that exhibit positive spatial dependence, we can usually reduce the model dimension by at least half {\em a priori}. And \citet{sglmm} showed that a much greater reduction is possible in practice, with 50--100 eigenvectors being sufficient for most datasets. Moreover, a simple spherical Gaussian proposal distribution for $\bEta$ performs well because the elements of $\bEta$ are approximately {\em a posteriori} uncorrelated owing to the orthogonality of the Moran basis.

Although using a truncated Moran basis dramatically reduces the time required to draw samples from the posterior, and the space required to store those samples, this approach does incur the substantial up-front burden of computing and eigendecomposing $M_x$. The efficiency of the former can be increased by storing $\ma$ in a sparse format \citep{Furrer:Sain:2010:JSSOBK:v36i10} and parallelizing the matrix multiplications. And we can more efficiently obtain the desired basis vectors by computing only the first $q$ eigenvectors of $M_x$ instead of doing the full eigendecomposition. This can be done using the Spectra library \citep{spectra}, for example.

We note that \citet{guanrsr} recently developed an approach to RSR for point-level data. Their approach is based on random projections \citep{sarlos2006improved,halko2011finding,banerjee2013efficient}.

\subsection{Computing for the SCRM}
\label{copcompute}

The hierarchical copula model and the direct copula model pose rather different computing challenges. And that is not the only important difference between the two models. A sufficiently substantive discussion of this issue is beyond the scope of this article, but it is worth mentioning that the hierarchical SCRM may be more appealing from a modeling point of view \citep{musgrove2016hierarchical} but suffers from certain limitations when employed as a data-analytic tool \citep{han2016correlation}. For this reason we will focus on copCAR \citep{hughes2014copcar}, a form of the direct copula model, here and in Section~\ref{simstudy}.

copCAR employs the CAR copula, a Gaussian copula (or other suitable copula) based on the proper CAR described above. Recall that the proper CAR has precision matrix $\tau\mq$, where $\mq=\diag(\ma\bone)-\rho\ma$. Since a copula is scale free, we do not need $\tau$, but omitting $\tau$ does not leave us with an inverse correlation matrix because the variances $\bsigma^2=(\sigma_1^2,\dots,\sigma_n^2)^\prime=\vdiag(\mq^{-1})$ are not equal to 1. We could rescale $\mq$ so that its inverse is a correlation matrix, i.e., we could construct a Gaussian copula using $\mlam^{1/2}\mq\mlam^{1/2}$, where $\mlam=\diag(\bsigma^2)$. In fact, rescaling is necessary in the general case lest the model be unidentifiable with respect to the variances. For copCAR, however, rescaling is unnecessary because the variances $\bsigma^2$ are not free parameters; the variances are entirely determined by $\mq$'s only dependence parameter, $\rho$, which is not a scale parameter. Since using $\mq$ itself leads to an identifiable model, rescaling would merely slow computation. Thus copCAR employs the CAR correlation structure indirectly, by using $\mq$ along with the variances $\bsigma^2$. This leads to the CAR copula:
\begin{align}
\label{carcop}
\Phi_{\bzero,\mq^{-1}}\{\Phi_{\sigma_1}^{-1}(u_1),\dots,\Phi_{\sigma_n}^{-1}(u_n)\},
\end{align}
where $\Phi_{\sigma_i}$ denotes the distribution function of the normal distribution with mean 0 and variance $\sigma_i^2$.

The model specification can be completed by pairing the CAR copula with a set of suitable marginal distributions for the outcomes. The copula and the marginals are linked by way of the probability integral transform. Specifically, if $\bZ=(Z_1,\dots,Z_n)^\prime$ are the observations, and  $F_1,\dots,F_n$ are the desired marginal distribution functions, we have $Z_i=F_i^{-1}(U_i)$, where $\bU=(U_1,\dots,U_n)'$ is a realization of the copula. We will assume Bernoulli marginal distributions with expectations $\{1+\exp(-\bx_i^\prime\bbeta)\}^{-1}$.

Unless $n$ is quite small, computation of the copCAR likelihood is infeasible (when the marginals are discrete) because the multinormal cdf is unstable in high dimensions and because the likelihood contains a sum of $2^n$ terms. For Bernoulli marginals, a composite marginal likelihood approach \citep{varin2008composite} performs well. The objective function is a product of pairwise likelihoods:
\begin{align*}
L_\cml(\bs{\theta}\mid\bZ) &= \mathop{\prod_{i,j\in\{1,\dots,n\}}}_{i\neq j}\;\sum_{j_1=0}^1\sum_{j_2=0}^1(-1)^kH_{ij}(U_{ij_1},U_{jj_2}),
\end{align*}
where $H_{ij}$ denotes the bivariate Gaussian copula with covariance matrix
\[
\mv^{ij}=\begin{pmatrix}\sigma_i^2&(\mq^{-1})_{ij}\\(\mq^{-1})_{ij}&\sigma_j^2\end{pmatrix}.
\]
This implies the log composite likelihood
\begin{align}
\label{likecl}
\ell_\cml(\bs{\theta}\mid\bZ) &= \mathop{\sum_{i\in\{1,\dots,n-1\}}}_{j\in\{i+1,\dots,n\}}\log\left\{\sum_{j_1=0}^1\sum_{j_2=0}^1(-1)^k\Phi_{\bzero,\mv^{ij}}(Y_{ij_1},Y_{jj_2})\right\},
\end{align}
where $Y_{\bullet 0}=\Phi_{\sigma_\bullet}^{-1}\{F_\bullet(Z_\bullet)\}$ and $Y_{\bullet 1}=\Phi_{\sigma_\bullet}^{-1}\{F_\bullet(Z_\bullet-1)\}$. Optimization of (\ref{likecl}) yields $\hat{\bs{\theta}}_\cml$. 

While $\hat{\bbeta}_\cml$ tends to be approximately normally distributed, $\hat{\rho}_\cml$ tends to be left skewed when $\rho$ is close to 1. This implies that asymptotic inference for $\rho$ tends to result in poor coverage rates. This can be avoided by using a parametric bootstrap, but a parametric bootstrap is rather burdensome computationally. Luckily, a simple reparameterization yields an approximately normally distributed estimator because the objective function for the reparameterized model is approximately quadratic with constant Hessian \citep{Geyer2005Le-Cam-Made-Sim}. Specifically, for $\btheta=(\bbeta^\prime,\Phi^{-1}(\rho))^\prime$, we have
\begin{align*}
\sqrt{n}(\hat{\btheta}_\cml-\btheta) &\;\;\;\Rightarrow\;\;\; \nrm\{\bzero,\;\info_\cml^{-1}(\btheta)\meat_\cml(\btheta)\info_\cml^{-1}(\btheta)\},
\end{align*}
where $\info_\cml$ is the Fisher information matrix and $\meat_\cml$ is the variance of the score:
\[
\meat_\cml(\btheta)=\var\nabla\ell_\cml(\btheta\mid\bZ).
\]
Note that the asymptotic covariance matrix for the CML estimator is a Godambe information matrix \citep{godambe1960optimum} because $\ell_\cml$ is misspecified. The matrix can be estimated in the same manner as we described above for the autologistic model.

The form of $\ell_\cml$ given in (\ref{likecl}) requires four evaluations of the bivariate normal cdf for each of the $n(n-1)/2$ pairs of observations. This computation is rather expensive even for fairly small samples.

In a spatial setting we can expect a pair of nearby observations to carry more information about dependence than a pair of more distant observations. Others have found, in a variety of contexts, that retaining the contributions to the CML made by more distant pairs of observations decreases not only the computational efficiency of the procedure but also the statistical efficiency of the estimator \citep{varin2009pairwise,apanasovich2008aberrant}. Hence, we allow only pairs of adjacent observations to contribute to the copCAR CML. This means replacing (\ref{likecl}) with
\begin{align}
\label{newcml}
\ell_\cml(\bs{\theta}\mid\bZ) &= \mathop{\sum_{i,j:\,(i,j)\in E}}_{i<j}\log\left\{\sum_{j_1=0}^1\sum_{j_2=0}^1(-1)^k\Phi_{\bzero,\mv^{ij}}(Y_{ij_1},Y_{jj_2})\right\}.
\end{align}
If thoughtfully implemented, optimization of (\ref{newcml}) is efficient enough to permit analysis of larger areal datasets.

\section{Application of Various Spatial Regression Models to Simulated Binary Data}
\label{simstudy}

\begin{figure}[h]
   \centering
   \includegraphics[scale=.5]{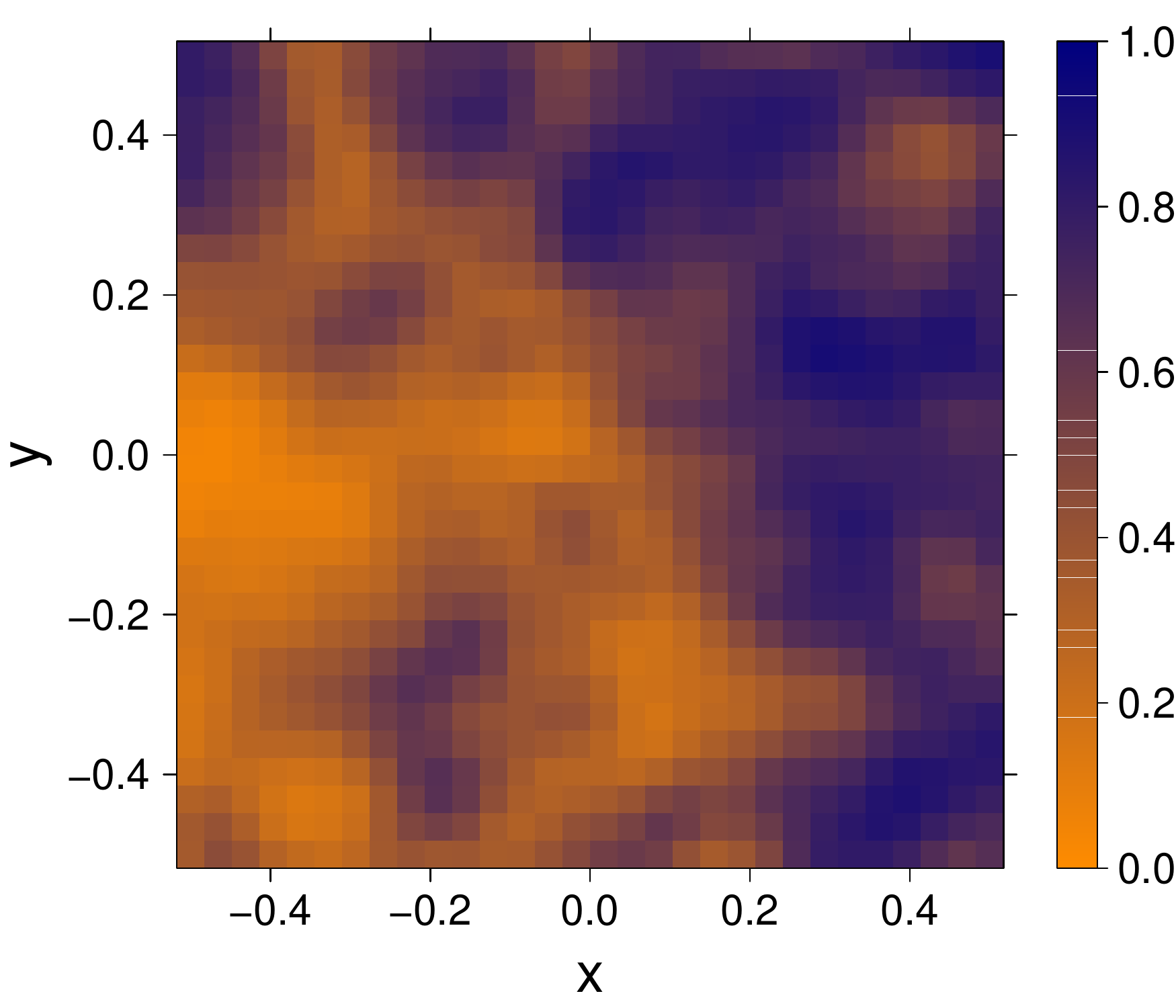}
   \caption{The mean surface for the simulation study.}
   \label{pfig}
\end{figure}

Our simulation study focused on binary areal outcomes, for the reasons given above. We simulated those outcomes on  the $30\times 30$ square lattice. This data size kept the computational burden manageable while giving all of the approaches a fighting chance at performing well. Our mean surface was a function of the $x$ and $y$ coordinates of the lattice points, $\bx=(x_1,\dots,x_n)'$ and $\by=(y_1,\dots,y_n)'$, respectively, which we restricted to the unit square centered at the origin. While simulating data we used linear predictor $\beta_0+\beta_1\bx_1+\beta_2\bx_2$, where $\bx_1=\bx$ and $\bx_2=\bx+\by+3\loc$. Vector $\loc$ exhibits a smaller-scale spatial pattern than do $\bx$ and $\by$; this lends more interesting spatial structure to the mean surface and ensures that $\bx_1$ and $\bx_2$ are substantially, but not strongly, correlated ($\text{cor}(\bx_1,\bx_2)=0.45$ rather than 0.71). We let $\bbeta=(0.2, 1, 1)'$, which implies a mean vector equal to
\[
\bp=\{\bone+\exp(-0.2-\bx_1-\bx_2)\}^{-1}=\{\bone+\exp(-0.2-\bx-\bx-\by-3\loc)\}^{-1}.
\]
These means are shown in Figure~\ref{pfig}.

Predictor $\bx_2$ was our unmeasured confounder and source of extra-$\mx\bbeta$ spatial variation. That is, we analyzed the data using $\mx=(\bone\;\bx_1)$, which implies that $\umx=\bx_2$. More specifically, to each of 100 simulated datasets we applied six models:
\begin{enumerate}
\item the ordinary logistic regression model with linear predictor $\beta_0+\beta_1\bx_1$;
\item the centered autologistic model having regression component $\beta_0+\beta_1\bx_1$;
\item the copCAR model with $\mathcal{B}er[\{\bone+\exp(-\beta_0-\beta_1\bx_1)\}^{-1}]$ marginals;
\item the traditional CAR model having regression component $\beta_0+\beta_1\bx_1$;
\item the sparse RSR model of \citeauthor{sglmm}, having regression component $\beta_0+\beta_1\bx_1$ and using the first $q=100$ eigenvectors of $M_x$, where $\mx=(\bone\;\bx_1)$; and
\item the sparse RSR model of \citeauthor{sglmm} along with the posterior predictive approach of \citeauthor{hanks}
\end{enumerate}
The results are provided in Table~\ref{simtab1}. We see that the RSR approach of \citeauthor{sglmm} performed better than the other approaches. The RSR estimator of $\beta_1$ has the smallest bias and mean squared error, and strikes the best balance between coverage rate and type II error rate. The RSR model also offers the most accurate prediction. The traditional CAR model, along with the RSR approach of \citeauthor{hanks}, resulted in very high coverage rates at the cost of very high type II error rates. The other three models performed poorly with respect to coverage rate and prediction.

\begin{table}[h]
   \centering
   \begin{scriptsize}
   \renewcommand{\arraystretch}{1.7}
   \begin{tabular}{rccccc}
      \hline
      Model & \parbox{1.4cm}{\vspace{1ex}\centering Med. Est.\\ of $\beta_1=1$} & \parbox{1.5cm}{\vspace{1ex}\centering Med. CI\\ Width} & MSE & Coverage Rate $-$ Type II Rate & Med. $\Vert\hat{\bp}-\bp\Vert$\vspace{1ex}\\\hline\hline
      Ordinary Logistic & 2.11  & 0.97 & 1.29  & \phantom{00}0\% $-$ \phantom{0}0\% = \phantom{0}0 & 4.93\\
      Centered Autologistic & 2.17  & 1.17 & 1.44  & \phantom{00}0\%  $-$ \phantom{0}0\% = \phantom{0}0  & 4.18\\
      copCAR & 2.15  & 1.26 & 1.36  & \phantom{00}0\%  $-$ \phantom{0}0\% = \phantom{0}0  & 4.93\\
      Traditional CAR & 2.35  & 5.27 & 2.59  & \phantom{0}99\%  $-$ 61\% = 38  & 3.21\\
      RSR ($q=100$) & \framebox{2.01}  & 2.30 & \framebox{1.18} & \framebox{\phantom{0}56\%  $-$ \phantom{0}2\% = 54} & \framebox{3.01}\\
      Adjusted RSR ($q=100$) & 2.01  & 5.75 &  3.51 & 100\%  $-$ 91\% = \phantom{0}9 & 3.01\\
      \hline
   \end{tabular}
   \end{scriptsize}
   \caption{Various performance measures for the first simulation study: median estimate of $\beta_1=1$, median 95\% confidence/credible region width, mean squared error, coverage rate minus type II error rate, and median prediction error.}
   \label{simtab1}
\end{table}

\begin{figure}[h]
   \centering
   \begin{tabular}{cc}
   $\bp$ & $\hat{\bp}$ RSR ($q=100$)\\
   \includegraphics[scale=.35]{p} & \includegraphics[scale=.35]{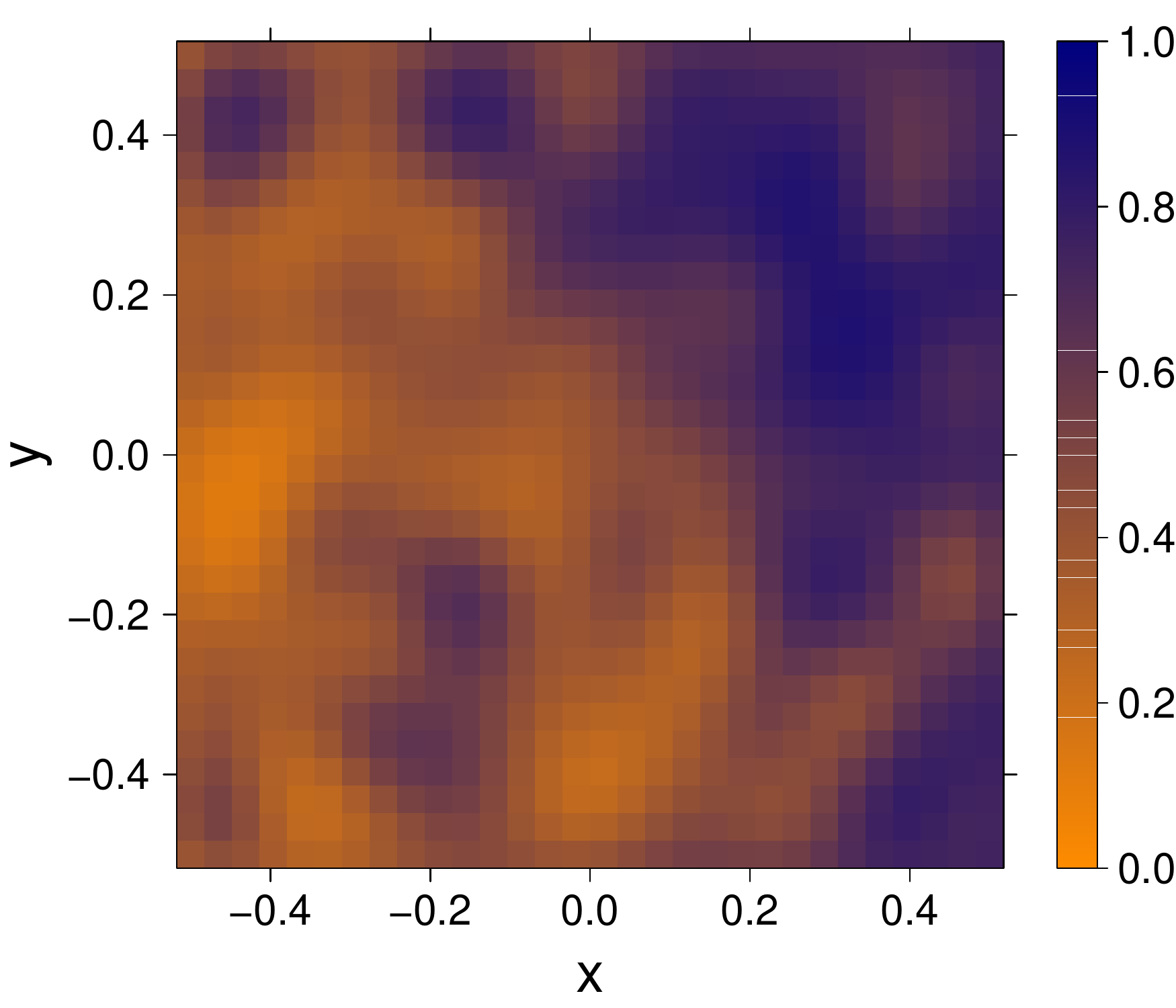}\vspace{1ex}\\
    $\hat{\bp}$ CAR & $\hat{\bp}$ Autologistic\\
    \includegraphics[scale=.35]{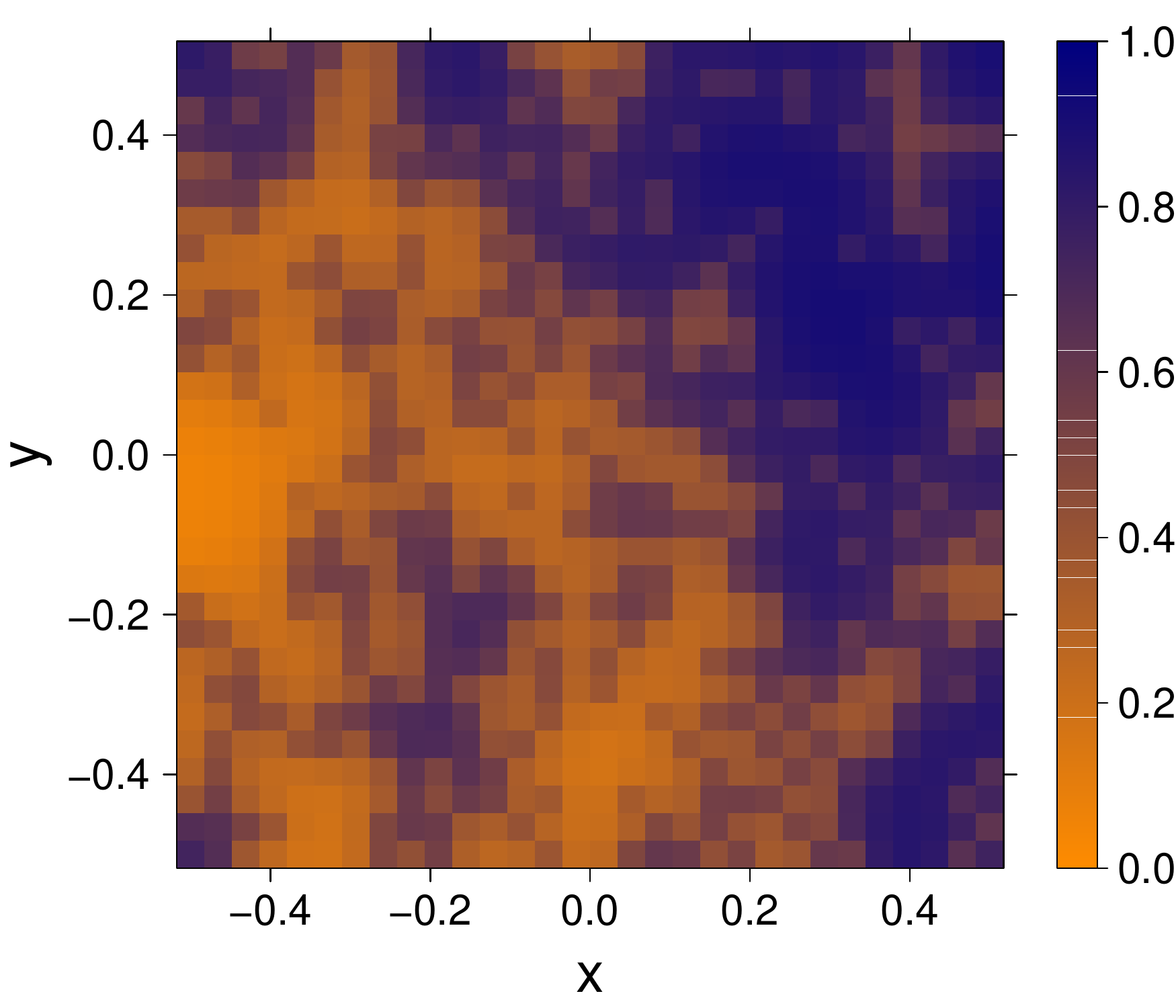} & \includegraphics[scale=.35]{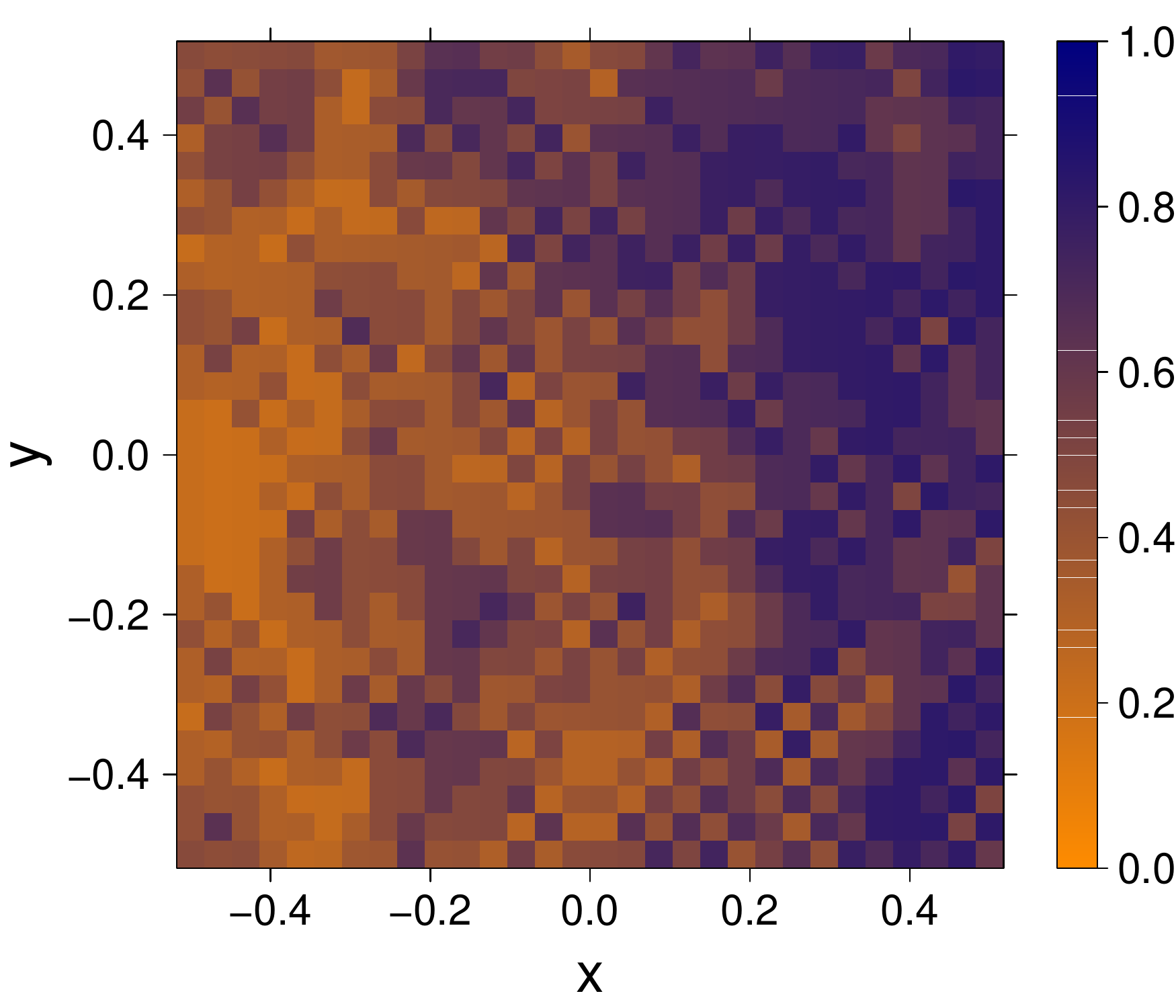}
   \end{tabular}
   \caption{Predictions for a single simulated dataset. The truth is shown in the upper left panel.}
   \label{predfig}
\end{figure}

Predictions for a single dataset are shown in Figure~\ref{predfig}. The autologistic model and the CAR model clearly undersmooth. The CAR model's undersmoothing is less dramatic, but it is perhaps surprising that the CAR model undersmoothes at all given that it has $n$ spatial random effects. (Note that we could force $\hat{\bpsi}$ to be smoother by using $\mq^k\;(k\geq 2)$ in place of $\mq$ \citep{GMRFbook}.)

\section{Bayesian Spatial Filtering}
\label{bayesfilter}

In this section we will develop, and assess the performance of, Bayesian spatial filtering, which possesses the computational advantages and good predictive performance of RSR while allowing for some advantages in regression inference. We begin by describing classical spatial filtering.

\subsection{Classical Spatial Filtering}
\label{filter}

In developing the SAMM, \citet{sglmm} drew inspiration from spatial filtering \citep{Griffith2003Spatial-Autocor}, which uses a basis expansion to accommodate any extra-$\mx\bbeta$ spatial pattern exhibited by the response vector, resulting in conventional residuals, i.e., residuals having at most trace spatial dependence. This implies that spatial filtering can reveal extra-$\mx\bbeta$ structure  while permitting the analyst to apply ordinary, well-understood diagnostic techniques to the residuals.

The basis used most often in spatial filtering are eigenvectors of the Moran operator for $\bone$: $M_1=(\id-n^{-1}\bone\bone')\ma(\id-n^{-1}\bone\bone')$. This yields vectors that reside in $C(\bone)^\perp$. (Recall that the SAMM employs basis vectors from $C(\mx)^\perp$, where $\mx$ typically contains $\bone$ along with one or more spatially structured predictors.) Considerable dimension reduction can be achieved by using only $q\ll n$ basis vectors. If we store said vectors as the columns of matrix $\mF_{n\times q}$, say, the filtering linear predictor can be written as
\begin{align*}
\vg(\bmu) &= \mx\bbeta + \mF\bEta,
\end{align*}
where $\bEta$ is once again a $q$-vector of coefficients.

Since constructing $\mF$ requires that $M_1$ be computed and eigendecomposed, it is clear that spatial filtering and the SAMM have much in common from a computational point of view. There are two key differences, however. First, to our knowledge, there are no Bayesian approaches for spatial filtering; practitioners estimate $\bEta$ by optimizing a likelihood or a composite likelihood. The choice of objective function of course has a substantial impact on computational complexity. And second, the columns of $\mF$ can be chosen using any of a number of methods (three will be discussed shortly). Those methods vary greatly in sophistication and computational complexity. It is not clear how they compare to one another with respect to quality of regression inference or quality of prediction, however.

\citet{chun2016eigenvector} recommend that the first
\[
q_0 = \frac{n_\texttt{+}}{1 + \exp[2.148-\{6.1808\,(z_\mi+0.6)^{0.1742}\}/n_\texttt{+}^{0.1298}+3.3534/(z_\mi+0.6)^{0.1742}]},
\]
eigenvectors be included initially in a stepwise, ordinary GLM analysis with a significance level of 0.2. Here, $n_\texttt{+}$ is the number of positive eigenvalues of $M_1$ and $z_\mi$ is the $z$ score of Moran's $I$ for the response.

For a binary response, another possibility is to do a two-sample $t$ test for each of the first few hundred eigenvectors, where the eigenvector of interest is treated as the response and $\bZ$ is used as the grouping variable. Any eigenvector that yields a $p$-value smaller than 0.1, say, is then included in the analysis. In this scheme, the number of variables may or may not be further reduced using a stepwise procedure.

A third approach to spatial filtering is to include Moran eigenvectors in a spatial model and use that model's dependence component to decide which eigenvectors to retain. For example, one might use some procedure to choose the Moran eigenvectors that lead to $\hat{\kappa}\approx 0$ for an appropriate automodel, or $\hat{\rho}\approx 0$ for a model that employs the proper CAR. It is this technique for which spatial filtering is named, since here an explicit aim is to remove (filter) spatial dependence from the response \citep{Griffith:2004p814}.

\subsection{A Bayesian Approach to Spatial Filtering}
\label{bfilter}

We can develop a Bayesian approach to spatial filtering by replacing $\mm$ in the SAMM specification with a filtering design matrix $\mF$. Specifically, the Bayesian spatial filtering (BSF) model has the same transformed conditional mean as the classical spatial filtering model, namely,
\begin{align*}
\vg(\bmu) &= \mx\bbeta + \mF\bEta,
\end{align*}
where $\mF_{n\times q}$ contains the $q$ principle eigenvectors of $M_1$, and $\bEta$ is a $q$-vector of coefficients. Borrowing from the SAMM, the prior distribution for $\bEta$ is
\begin{align}
\label{etaprior}
\bEta\sim\nrm\{\bzero,(\tau\mF'\mq\mF)^{-1}\},
\end{align}
where $\mq$ is once again the Laplacian of $G$. The BSF model, like the SAMM, assigns $\bbeta$ a spherical Gaussian prior with a large variance, and assigns the smoothing parameter $\tau$ a gamma prior with shape parameters 0.5 and 2,000. Note that the latter prior, having a large mean, discourages artifactual spatial structure in the posterior \citep{Kelsall1999Discussion-of-B}.

\begin{figure}[h]
   \centering
   \includegraphics[scale=.4]{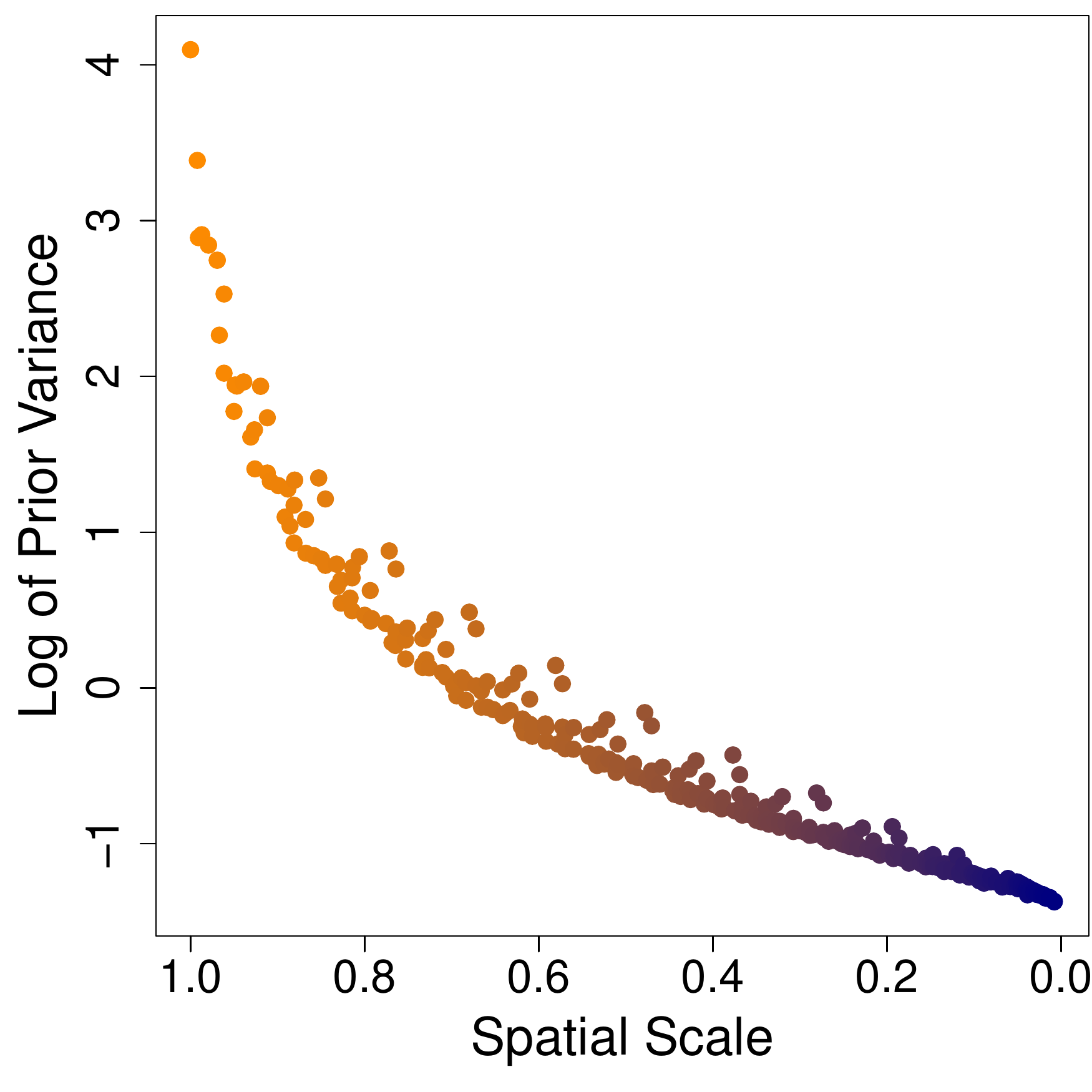}
   \caption{Prior variances (on the log scale and for $\tau=1$) for the elements of $\bEta\sim\nrm\{\bzero,(\tau\mF'\mq\mF)^{-1}\}$ from the second simulation study. The variances decrease rapidly as the spatial scale decreases, which prevents overfitting. }
   \label{varfig}
\end{figure}

Since $\bEta$ are regression coefficients, one may be tempted to assign $\bEta$ a spherical Gaussian prior instead of the above mentioned prior. This would be a mistake, however, for (\ref{etaprior}) is not arbitrary (see \citet{Reich:2006p1157} and/or \citet{sglmm} for derivations) but is, in fact, very well suited to the task at hand. Specifically, two characteristics of (\ref{etaprior})---along with the above mentioned prior for $\tau$---discourage overfitting even when $q$ is too large for the dataset being analyzed. First, the prior variances are commensurate with the spatial scales of the predictors in $\mF$ (Figure~\ref{varfig}). This shrinks toward zero the coefficients corresponding to predictors that exhibit small-scale spatial variation. Additionally, the correlation structure of (\ref{etaprior}) effectively reduces the degrees of freedom in the smoothing component of the model.

If the response is non-Gaussian, $\bbeta$ and $\bEta$ are updated using Metropolis--Hastings random walks with Gaussian proposals. The proposal covariance matrix for $\bbeta$ is the estimated asymptotic covariance matrix from an ordinary GLM fit to the data, which generally yields an acceptance rate around 50\%. The proposal for $\bEta$ is spherical Gaussian---with standard deviation $\sigma_\eta$, say. A sensible default value for $\sigma_\eta$ is 0.1, but a smaller value may be required to achieve large enough acceptance rates for larger datasets. The update for $\tau$ is a Gibbs update irrespective of the response distribution. If the response is Gaussian distributed, all updates are Gibbs updates. Note that the BSF MCMC sampler is very easy to tune since $\sigma_\eta$ is the only tuning parameter (unless the outcomes are Gaussian, in which case no tuning is required).

Bayesian spatial filtering for point-level outcomes can be accomplished analogously by adapting \citeapos{guanrsr} random-projection framework. The resulting BSF model for point-level data is different from the areal BSF model in a potentially important way, however. Since \citeauthor{guanrsr} obtain their basis vectors by eigendecomposing $\msigma$, they, quite naturally, assign a spherical Gaussian prior to $\bEta$. Because a spherical Gaussian prior lacks the appealing attributes of (\ref{etaprior}), \citeauthor{guanrsr} recommend that $q=\rank(\mF)$ be chosen in a pre-processing step. It is not clear how well this approach performs compared to the use of a prior similar to (\ref{etaprior}).

\subsection{Application of BSF to Simulated Binary Data}
\label{bfiltersim}

As a followup to the simulation study described in Section~\ref{simstudy}, we applied our BSF model to the simulated datasets. We used four different values for $q=\rank(\mF)$: 50, 100, 200, and 400. The results are given in Table~\ref{simtab2}. 

For smaller values of $q$, $\bhat_\bsf$ has smaller bias and MSE than any of the other estimators considered here, and yields a higher coverage rate while keeping the type II rate very low. As $q$ becomes large, the bias and MSE of $\bhat_\bsf$ grow. Eventually the coverage rate begins to decrease, the type II rate to increase. The BSF model also performs very well at prediction.

The BSF model accomplishes all of this through judicious use of multicollinearity. Recall that the basis vectors used in RSR are (at least nearly) uncorrelated with the columns of $\mx$. This is not the case for the BSF model since spatial filtering employs eigenvectors of $M_1$. As we increase $q$, we introduce more and more multicollinearity between $\mx$ and $\mF$. Up to a point, this alleviates unmeasured confounding to some extent (hence the reduced bias), and adjusts the variance upward by a modest amount (hence the increased coverage rate). As $q$ gets large, the linear predictor becomes rather redundant, causing the BSF model to perform much like the CAR model (increased bias and type II error rate).

\begin{table}[h]
   \centering
   \begin{scriptsize}
   \renewcommand{\arraystretch}{1.7}
   \begin{tabular}{rccccc}
      \hline
      Model & \parbox{1.5cm}{\vspace{1ex}\centering Med. Est.\\ of $\beta_1=1$} & \parbox{1.5cm}{\vspace{1ex}\centering Med. CI\\ Width} & MSE & Coverage Rate $-$ Type II Rate & Med. $\Vert\hat{\bp}-\bp\Vert$\vspace{1ex}\\\hline\hline
      Ordinary Logistic & 2.11  & 0.97 & 1.29 & \phantom{00}0\% $-$ \phantom{0}0\% = \phantom{0}0 & 4.93 \\
      Centered Autologistic & 2.17  & 1.17 & 1.44  & \phantom{00}0\%  $-$ \phantom{0}0\% = \phantom{0}0  & 4.18\\
      copCAR & 2.15  & 1.26 & 1.36  & \phantom{00}0\%  $-$ \phantom{0}0\% = \phantom{0}0  & 4.93\\
      Traditional CAR & 2.35  & 5.27 & 2.59  & \phantom{0}99\%  $-$ 61\% = 38  & 3.21\\
      RSR ($q=100$) & 2.01  & 2.30 & 1.18  & \phantom{0}56\%  $-$ \phantom{0}2\% = 54 & 3.01\\
      Adjusted RSR ($q=100$) & 2.01  & 5.75 &  3.51 & 100\%  $-$ 91\% = \phantom{0}9 & 3.01\\
      BSF ($q=50$) & \framebox{1.87}  & 2.08 & \framebox{0.94}  & \framebox{\phantom{0}63\% $-$ \phantom{0}2\% = 61}   & \framebox{3.07}\\
      BSF ($q=100$) & \framebox{1.89}  & 2.34 &  \framebox{0.96} & \framebox{\phantom{0}74\% $-$ \phantom{0}2\% = 72} & \framebox{3.01}\\
      BSF ($q=200$) &  1.99 & 2.69 & 1.25  & \phantom{0}81\% $-$ \phantom{0}2\% = 79 & 2.99\\
      BSF ($q=400$) & 2.25  & 3.23 & 1.88  & \phantom{0}67\% $-$ 20\% = 47 & 3.01\\
      \hline
   \end{tabular}
   \end{scriptsize}
   \caption{The performance of Bayesian spatial filtering.}
   \label{simtab2}
\end{table}





\section{Conclusion}
\label{conclusion}

When unmeasured confounding is the source of extra-$\mx\bbeta$ spatial variation in a response variable, spatial regression models struggle to perform well. In Section~\ref{simstudy} we saw that the autologistic model and the copCAR model (examples of the automodel and the spatial copula regression model, respectively) perform rather poorly, about as poorly as a non-spatial model. Spatial mixed-effects models perform better but still have weaknesses. The traditional SGLMM, for example, is badly spatially confounded and computationally burdensome, and often undersmoothes. Restricted spatial regression offers an appealing alternative, for RSR reduces bias and mean squared error, provides a more sensible balance between coverage rate and type II rate, smoothes very effectively, and permits efficient computation. Yet there is room for improvement.

In the latter part of this article we developed Bayesian spatial filtering, which performs as well as RSR with respect to prediction and computational complexity while besting RSR in terms of bias, mean squared error, and coverage rate. BSF does this by using an expansion in a well-chosen basis to introduce an appropriate amount of spatial confounding. This situates the BSF model on a continuum of spatial confounding, with the non-spatial model and the CAR model at either end:
\[
(\text{Non-Spatial Model})\;\;\;\underset{q\searrow 0}{\Longleftarrow}\;\;\;(\text{BSF Model})\;\;\;\underset{q\nearrow n}{\Longrightarrow}\;\;\;(\text{CAR Model}).
\]

\bibliography{refs}
\bibliographystyle{apa}

\end{document}